\documentclass[a4paper,11pt]{article}
\pdfoutput=1 

\usepackage{jinstpub} 

\usepackage{lineno}
\usepackage[numbers]{natbib}
\usepackage{graphicx}
\usepackage{subcaption}
\usepackage{multirow}
\usepackage{multicol}
\usepackage[table]{xcolor}
\usepackage{amsmath}
\usepackage{booktabs}
\usepackage{wasysym}
\usepackage{url}
\usepackage{xcolor}

\newcommand{\NUCLEUS}[1]{\textsc{Nucleus} }
\newcommand{\cevns}[1]{CE$\nu$NS}

\title{Development of a compact muon veto for the \NUCLEUS{}experiment}

\author[a,b,1]{V.~Wagner\note{Corresponding author},}
\author[a]{R.~Rogly,}
\author[a,b]{A.~Erhart,}
\author[a]{V.~Savu,}
\author[a]{C.~Goupy,}
\author[a]{D.~Lhuillier,}
\author[a]{M.~Vivier,}
\author[b]{L.~Klinkenberg,}
\author[c]{G.~Angloher,}
\author[c,d]{A.~Bento,}
\author[c]{L.~Canonica,}
\author[e]{F.~Cappella,}
\author[e]{L.~Cardani,}
\author[e]{N.~Casali,}
\author[f,g]{R.~Cerulli,}
\author[e,h]{I.~Colantoni,}
\author[e]{A.~Cruciani,}
\author[e,i]{G.~del Castello,}
\author[j]{M.~Friedl,}
\author[c]{A.~Garai,}
\author[j]{V.M.~Ghete,}
\author[k,l]{V.~Guidi,}
\author[c]{D.~Hauff,}
\author[b]{M.~Kaznacheeva,}
\author[b]{A.~Kinast,}
\author[j]{H.~Kluck,}
\author[b]{A.~Langenk\"{a}mper,}
\author[a,m]{T.~Lasserre,}
\author[c]{M.~Mancuso,}
\author[a]{B.~Mauri,}
\author[l]{A.~Mazzolari,}
\author[a]{E.~Mazzucato,}
\author[a]{H.~Neyrial,}
\author[a]{C.~Nones,}
\author[b]{L.~Oberauer,}
\author[a,2]{A.~Onillon\note{Now at TU Munich},} 
\author[b]{T.~Ortmann,}
\author[b,n]{L.~Pattavina,}
\author[c]{F.~Petricca,}
\author[b]{W.~Potzel,}
\author[c]{F.~Pr\"{o}bst,}
\author[c]{F.~Pucci,}
\author[j,o]{F.~Reindl,}
\author[b]{J.~Rothe,}
\author[b]{N.~Schermer,}
\author[j,o]{J.~Schieck,}
\author[b]{S.~Sch\"{o}nert,}
\author[j,o]{C.~Schwertner,}
\author[a]{L.~Scola,}
\author[c]{L.~Stodolsky,}
\author[b]{R.~Strauss,}
\author[e]{C.~Tomei,}
\author[b]{K.~v.~Mirbach,}
\author[e,i]{M.~Vignati}
\author[b]{and A.~Wex}

\affiliation[a]{IRFU, CEA, Universit\'{e} Paris Saclay, F-91191 Gif-sur-Yvette, France}
\affiliation[b]{Physik-Department, Technische Universit\"at M\"unchen, D-85748 Garching, Germany}
\affiliation[c]{Max-Planck-Institut f\"ur Physik, D-80805 M\"unchen, Germany}
\affiliation[d]{CIUC, Departamento de Fisica, Universidade de Coimbra, P3004 516 Coimbra, Portugal}
\affiliation[e]{Istituto Nazionale di Fisica Nucleare -- Sezione di Roma, Roma I-00185, Italy}
\affiliation[f]{Istituto Nazionale di Fisica Nucleare -- Sezione di Roma "Tor Vergata", Roma I-00133, Italy}
\affiliation[g]{Dipartimento di Fisica, Universit\`{a} di Roma "Tor Vergata", Roma I-00133, Italy}
\affiliation[h]{Consiglio Nazionale delle Ricerche, Istituto di Nanotecnologia, Roma I-00185, Italy}
\affiliation[i]{Dipartimento di Fisica, Sapienza Universit\`{a} di Roma, Roma I-00185, Italy}
\affiliation[j]{Institut f\"ur Hochenergiephysik der \"Osterreichischen Akademie der Wissenschaften, A-1050 Wien, Austria}
\affiliation[k]{Dipartimento di Fisica, Universit\`{a} di Ferrara, I-44122 Ferrara, Italy}
\affiliation[l]{Istituto Nazionale di Fisica Nucleare -- Sezione di Ferrara, I-44122 Ferrara, Italy}
\affiliation[m]{APC, Universit\'{e} de Paris, CNRS, Astroparticule et Cosmologie, Paris F-75013, France}
\affiliation[n]{Istituto Nazionale di Fisica Nucleare -- Laboratori Nazionali del Gran Sasso, Assergi (L’Aquila) I-67100, Italy}
\affiliation[o]{Atominstitut, Technische Universit\"at Wien, A-1020 Wien, Austria}

\emailAdd{victoria.wagner@tum.de}

\abstract{
The \NUCLEUS{}experiment aims to measure coherent elastic neutrino nucleus scattering of reactor anti-neutrinos using cryogenic calorimeters. Operating at an overburden of 3~meters of water equivalent, muon-induced backgrounds are expected to be one of the dominant background contributions. Besides a high efficiency to identify muon events passing the experimental setup, the \NUCLEUS{}muon veto has to fulfill tight spatial requirements  to fit the constraints given by the experimental site and to minimize the induced detector dead-time. We developed highly efficient and compact muon veto modules based on plastic scintillators equipped with wavelength shifting fibers and silicon photo multipliers to collect and detect the scintillation light. In this paper, we present the full characterization of a prototype module with different light read-out configurations. We conclude that an efficient and compact muon veto system can be built for the \NUCLEUS{}experiment from a cube assembly of the developed modules. Simulations show that an efficiency for muon identification of $>99$ \% and an associated rate of $325$~Hz is achievable, matching the requirements of the \NUCLEUS{}experiment.
}


\arxivnumber{2202.03991} 

\collaboration[c]{\newline the \NUCLEUS{}collaboration}

\begin{document}
\maketitle

	\section{Introduction}
	\label{sec:intro}
	
	Coherent elastic neutrino-nucleus scattering (\cevns{})~\cite{PhysRevD.9.1389} can probe the Standard Model of particle physics in a way that is complementary to the low-energy tests performed to date and can be used to search for new physics~\cite{Scholberg:2005qs}.  
	Nuclear reactors are a favorable neutrino source due to their large flux of low energy anti-neutrinos ensuring a strong \cevns{} signal in the fully coherent regime. 
	Besides the advantage of studying \cevns{} of reactor anti-neutrinos, an experimental site close to a nuclear reactor poses several challenges: restricted access, limited space and shallow overburden as shielding against cosmic-ray induced backgrounds. 
	Thus, although the \cevns{} cross-section can be orders of magnitude larger than for inverse beta decay, efficient passive and active shieldings surrounding the target detectors are essential.
	
	Many \cevns{} experiments are currently operated or are being prepared at nuclear power and research reactors with varying overburden from 0 to 45~meters of water equivalent (m.w.e.)~\cite{Hakenmuller:2019ecb, SalagnacM7:2020, Aguilar:2019jlr, Singh:2017jow, Angloher:2019flc}. 
	At such conditions, the cosmic-ray induced background is dominated by muons and muon-induced neutrons produced in high-Z material, commonly used to shield external gamma rays~\cite{Heusser:1995wd}. 
	Muon-induced background is typically addressed two-fold: passive shieldings made e.g. from borated polyethylen (PE) are used to moderate and absorb neutrons, and active background rejection to discard prompt muon-associated events. For the latter, muons passing through the experimental setup need to be detected with a high efficiency.

	The \NUCLEUS{}experiment will operate cryogenic CaWO$_4$ and Al$_2$O$_3$ calorimeters at the Very-Near-Site (VNS) at the Chooz nuclear power plant in France~\cite{Angloher:2019flc}. 
	The VNS is a new experimental site located in an administrative building in-between the two 4.25~GW$_{\text{th}}$ reactor cores with an overburden of about 3~m.w.e. 
	To reach the benchmark background index of 100~counts/(keV$\cdot$kg$\cdot$d), muon-related backgrounds need to be reduced to a negligible contribution.
	Furthermore, due to the signal rise time of $\mathcal{O}$(100\,$\mu\text{s})$ of the \NUCLEUS{}target detectors, a muon identification rate of several hundred Hz implies a total dead time of a few percent~\cite{Angloher:2019flc}. 
	
	The \NUCLEUS{}muon veto will consist of several individual modules closely placed around the \NUCLEUS{}passive lead and PE  shielding. The modules themselves must exhibit a high efficiency to discriminate muon from ambient gamma events. 
	Ideally, the muon veto yields a $4\pi$ coverage of the lead shielding in order to minimize the probability to miss a muon passing through high-Z material and potentially producing secondary events. At the same time, the size of the overall muon veto needs to be kept at a minimum such that the muon identification rate and the associated dead-time stays moderate.  
	Thus, the individual panels need to be compact with a minimum of non-instrumented (dead) area and their placement around the passive shielding needs to be optimized. 
	
	In this paper we describe the full characterization of the prototype muon veto module made of BC-408 plastic scintillator read out with wavelength shifting (WLS) fibers coupled to silicon photo multipliers (SiPM). 
	Section~\ref{sec:Prototypes} presents the prototype module which holds different configurations of the WLS fibers and SiPM read-out. 
	The calibration of the SiPM is presented in section~\ref{sec:Calib}. 
	Section~\ref{sec:performance} focuses on the results of the characterization measurements: we compare the light yield (LY) and the muon-identification power of the different configurations. By combining measurements with GEANT4 Monte Carlo simulations, several effects which influence the LY homogeneity of the module can be disentangled.  
	We conclude with the presentation of the final \NUCLEUS{}muon veto shielding.  
	In  section~\ref{sec:GeoEff} we present geometrical simulations of the \NUCLEUS{}muon veto and estimate its (geometrical) veto efficiency together with the associated muon rate.

	\section{Prototype panel and test stand} 
	\label{sec:Prototypes}

\begin{figure}[h!]
\centering
   \begin{tabular}{m{0.66\textwidth} m{0.3\textwidth} }
     \begin{subfigure}[b]{0.66\textwidth}
        \includegraphics[width=\textwidth]{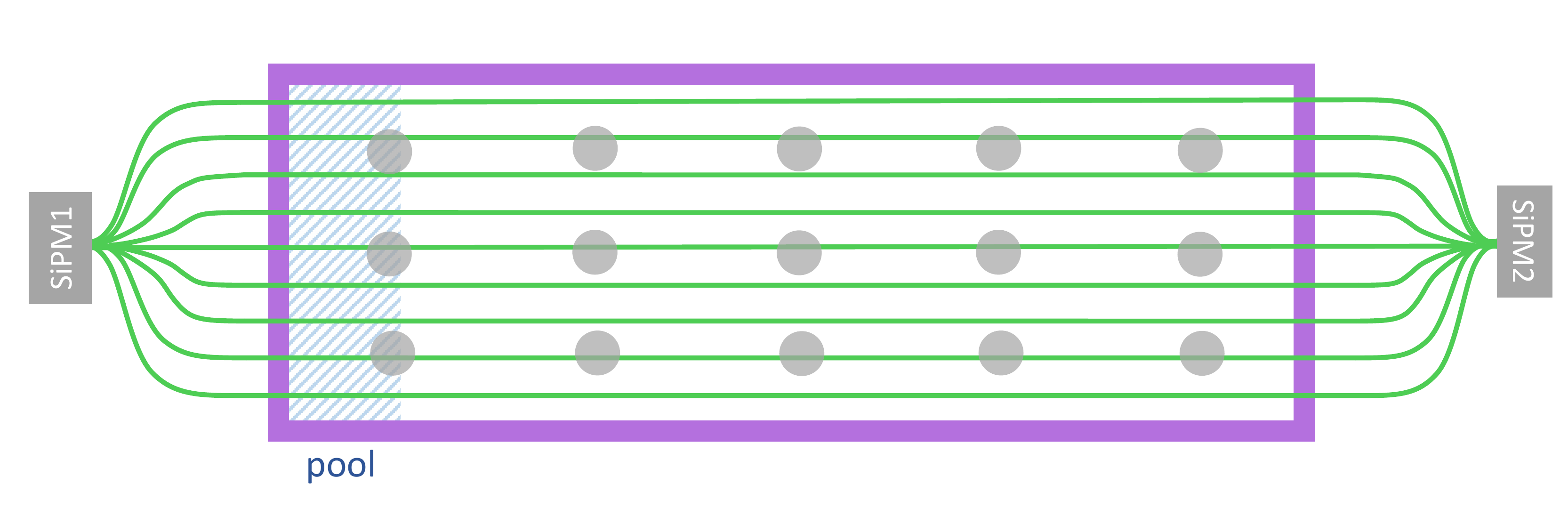}
            \caption{Straight fiber configuration (top side)}
            \label{fig:double_sided}
    \end{subfigure} 
    
      \begin{subfigure}[b]{0.66\textwidth}
        \includegraphics[width=\textwidth]{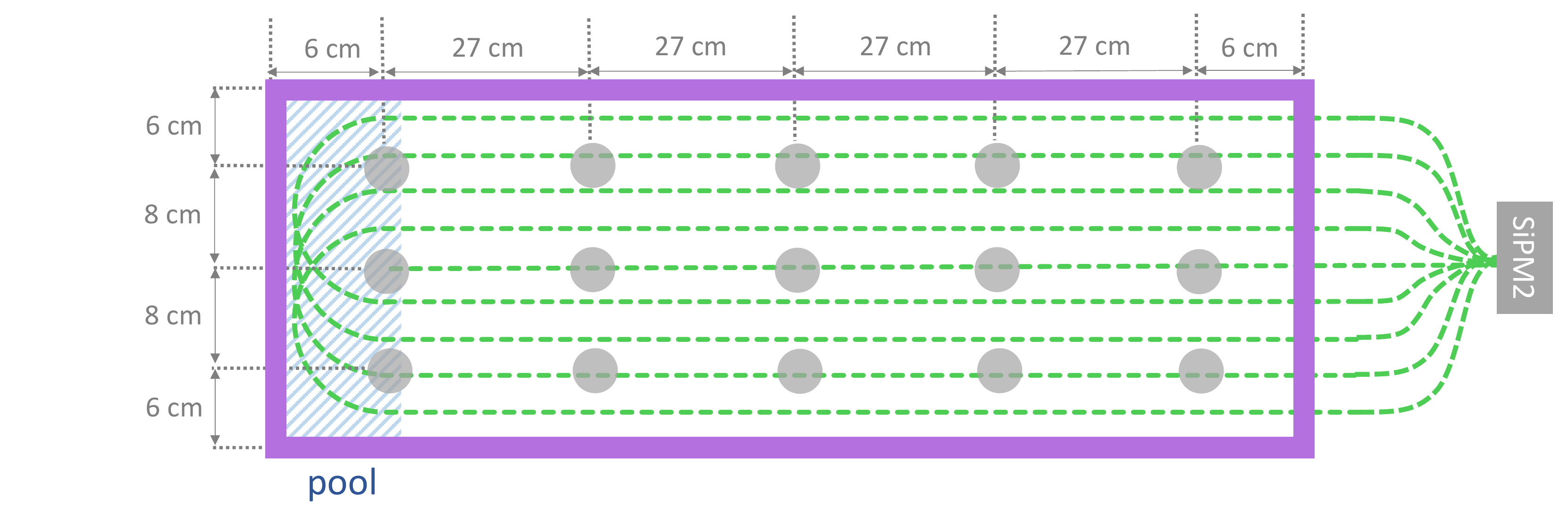}
            \caption{U-turn fiber configuration (bottom side)}
            \label{fig:U-turn}
    \end{subfigure} 
     & 
   \begin{subfigure}[b]{0.3\textwidth}
        \includegraphics[width=\textwidth]{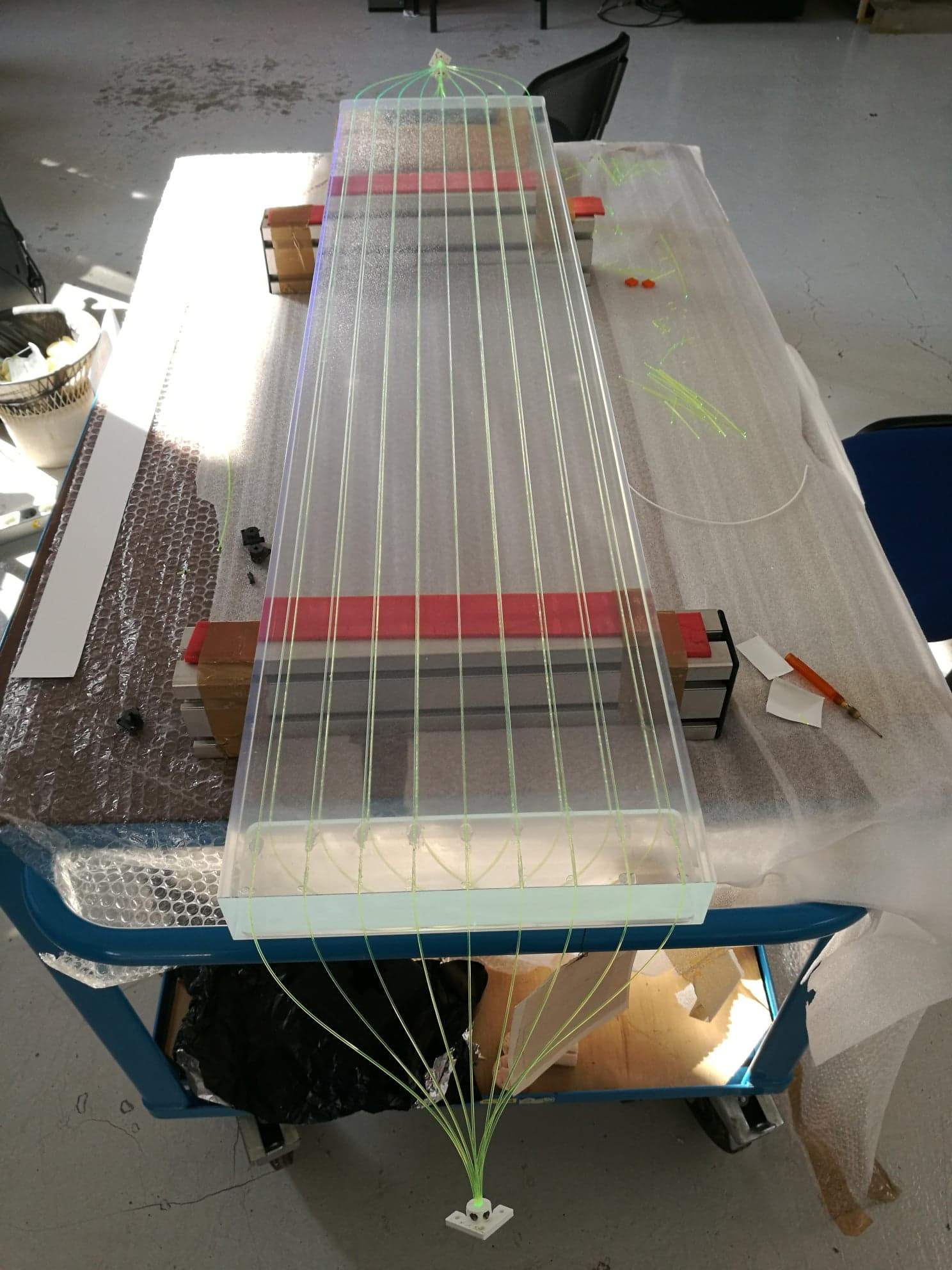}
   \caption{Prototype muon veto module}
    \label{fig:prototype} 
    \end{subfigure}

\\
    \end{tabular}
           \caption{Different fiber configurations: The straight fiber configuration (\subref{fig:double_sided}) with two SiPMs and the U-turn fiber configuration (\subref{fig:U-turn}) are shown. The grey circles indicate the measurement position discussed in section~\ref{sec:LYHomo}.  (\subref{fig:prototype}) shows a photograph of the \NUCLEUS{}muon veto prototype panel with the straight fiber configuration on the top side, and the U-turn configuration on the bottom. }
        \label{fig:configurations}
    \end{figure} 

Plastic scintillators coupled to SiPMs are known to be reliable radiation detectors. 
	At minimum ionization energy loss, a muon deposits about 2\,MeV/cm along its passage through a polyvinyltoluene-based plastic scintillator~\cite{Groom:2001kq}. 
	In order to distinguish muons from ambient $\gamma$-rays which extend up to 2.6\,MeV, we chose a baseline design with a 5\,cm thick plastic scintillator equipped with WLS fibers to guide the scintillation light towards a SiPM. 
	
	We chose the plastic scintillator BC-408 from Saint-Gobain~\cite{SaintGobain} since it features a large light output and large attenuation length, and is thus suitable for large area-modules. 
	The \NUCLEUS{}collaboration recovered 40 plastic scintillator panels from the CAMERA detector of the COMPASS-II experiment at the CERN SPS~\cite{Gautheron:2010wva} which was dismounted in 2018. 
    Each panel is about 1.6\,m long with a trapezoidal cross-section of 5\,cm height and a base of 30.3\,cm and 29.0\,cm. 
    As light sensor we chose the SiPM module PE3325-WB-TIA-SP from KETEK~\cite{KETEK}: a SiPM with a $3\times3~\text{mm}^2$ active area and a pixel size of $25\mu$m. The module features an integrated transimpedance amplifier.
    
    For a compact design, we chose to collect the light by WLS fibers rather than bulky light guides made e.g. from acrylic plastics. 
    The scintillation light produced in the plastic scintillator is absorbed by WLS fibers which run along the surface of the panel.
    WLS fibers feature an isotropic emission at a longer wavelength where part of the re-emitted light is transmitted by total internal reflection towards the SiPM which is mounted at one end of the fiber. 
    We chose commercially available multi-clad WLS fibers of type BC-91A from Saint-Gobain~\cite{SaintGobain2} with a cross-section of (1$\times$1)\,mm$^2$. The absorption and emission spectra match well the transmission spectrum of the BC-408 plastic scintillator and the range of high photo detection efficiency of the chosen SiPMs. 
    
    We made use of the high flexibility of the WLS fibers and tested three different configurations: (1) fibers in straight lines with light read-out at both fiber ends, figure~\ref{fig:double_sided}, and (2) fibers describing a U-turn and single SiPM read-out, see figure~\ref{fig:U-turn}. In a third (mirror) configuration (3), the second SiPM in figure~\ref{fig:double_sided} is replaced by a mirror at the open fiber end. 
    
    For the tests presented in this work, we built a prototype muon veto module made of a 1.2\,m long 28\,cm\,$\times$\,5\,cm rectangular-shaped BC-408 plastic scintillator, see figure~\ref{fig:prototype}. 
    In the case of the straight fiber configuration, figure~\ref{fig:double_sided}, the nine WLS fibers are running in 2\,mm deep grooves along the full length of the bottom plane of the plastic scintillator, each separated by a distance of 30\,mm. 
    For good optical coupling of the fibers to the scintillator, optical grease is inserted into the grooves. Tiny drops of glue at each end of the grooves keep the fibers in place. 
    At each end of the plastic scintillator, the fibers are bundled to a square of $(3\times3$)\,\text{mm}$^2$ by a 3-D printed plastic connector and attached to a SiPM module. The connector design ensures a precise alignment of the fiber cross-section with the active area of the SiPM as well as a safety gap of a few hundred microns between the two surfaces. The optical coupling is then achieved with optical grease. 
    The upper plane of the scintillator is used for the U-turn configuration. A ($275\times60\times5$)\,mm$^3$ pool is machined into the plastic scintillator at one end in addition to the nine straight grooves. This way, four fibers can describe a U-turn inside the pool and re-enter the groove with a bending radius of 6\,cm. This portion of the fibers stands in the air with no direct optical coupling to the scintillator. 
    A fifth straight fiber is placed in the middle groove, see figure~\ref{fig:U-turn}. 
    In order to avoid conflicts due to the crossing of the fibers at the level of the U-turns, the four pairs of grooves are machined with four different depths, ranging from 2\,mm to 5\,mm.
    
    To enhance light collection, the plastic scintillator is covered with a diffusive Lumirror E6SR foil from Toray~\cite{TORAY}. The full module including the SiPMs is enclosed in an aluminum light-tight box in which the SiPM modules are placed with a distance of 17\,cm with respect to the plastic scintillator in order to respect the WLS fibers' minimum recommended bending radius of 5\,cm. 
      
	The configuration with the straight fibers is the simplest design, expected to perform best in terms of light collection. However, this comes at the expense of a second SiPM read-out channel and additional un-instrumented (dead) volume inside the module box. 
	This dead volume is saved in the configurations where the WLS fibers run in a U-turn or where a mirror is placed at one end of the fibers. 
	For an efficient mirror configuration a highly reflective material which is carefully coupled to the open end of the fiber is required, adding complexity to the assembly. 
	The U-turn configuration combines compactness of the module and the advantage of a double sided read-out of the WLS fibers. However, the fibers are significantly longer with respect to the straight configurations, and light might be lost due to attenuation and the bending of the fibers. 

\begin{figure}[h!]
    \centering
    \includegraphics[width=0.75\textwidth]{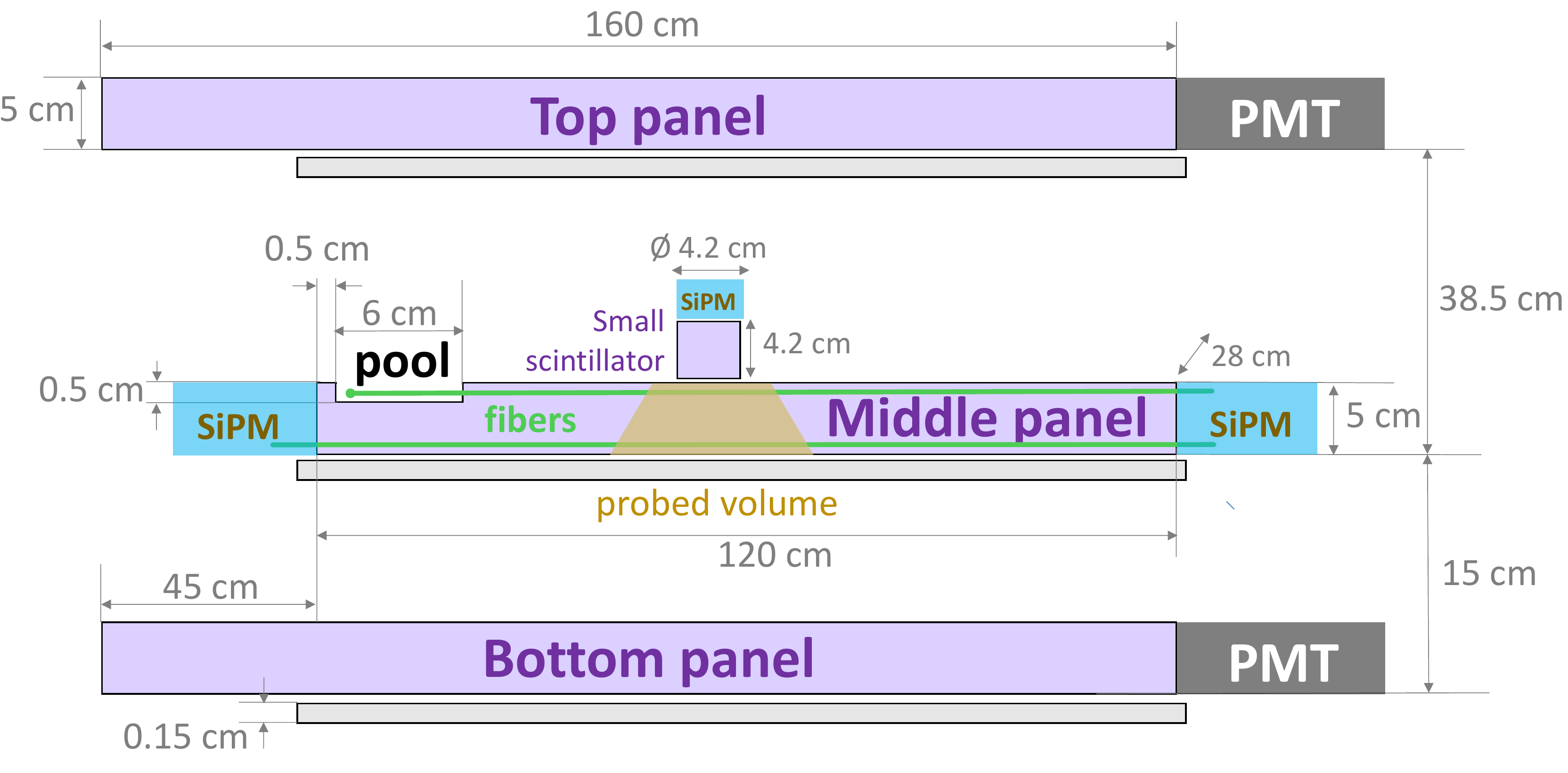}
    \caption{Muon veto test bench. Wood shelves are drawn in gray and plastic scintillators in purple. The middle panel is the \NUCLEUS{}muon veto prototype with two series of carved grooves: the top side carries the U-turn configuration with the pool and the bottom side the straight fiber configuration. In case of the mirror fiber configuration, one SiPM is replaced by a mirror. The blue area marks the non-instrumented part of the aluminium box where the SiPM module is placed.}
    \label{fig:Setup_scheme}
\end{figure}

 To characterize and compare the performance of the different fiber configurations, we built a dedicated test stand at CEA-Saclay. 
    Figure~\ref{fig:Setup_scheme} shows the sketch of the muon veto test bench. The prototype (middle panel) is placed in-between two 1.6\,m-long plastic scintillator panels each one coupled to a PMT. 
    These two panels (denoted top and bottom panel) are used to select muon events by means of a three fold coincidence in the two PMTs and the SiPM(s). The panels are placed at a distance of 38.5\,cm (top panel) and 15\,cm (bottom panel) with respect to the prototype. Thus, the angle of accepted muons passing through the prototype is restricted. 
    To select muons passing through a small volume of the prototype, an additional plastic scintillator with a diameter of 4.2\,cm and 4.2\,cm height read out with a SiPM is used. 
    
\begin{figure}[h!]
\centering
    \includegraphics[width=0.75\textwidth]{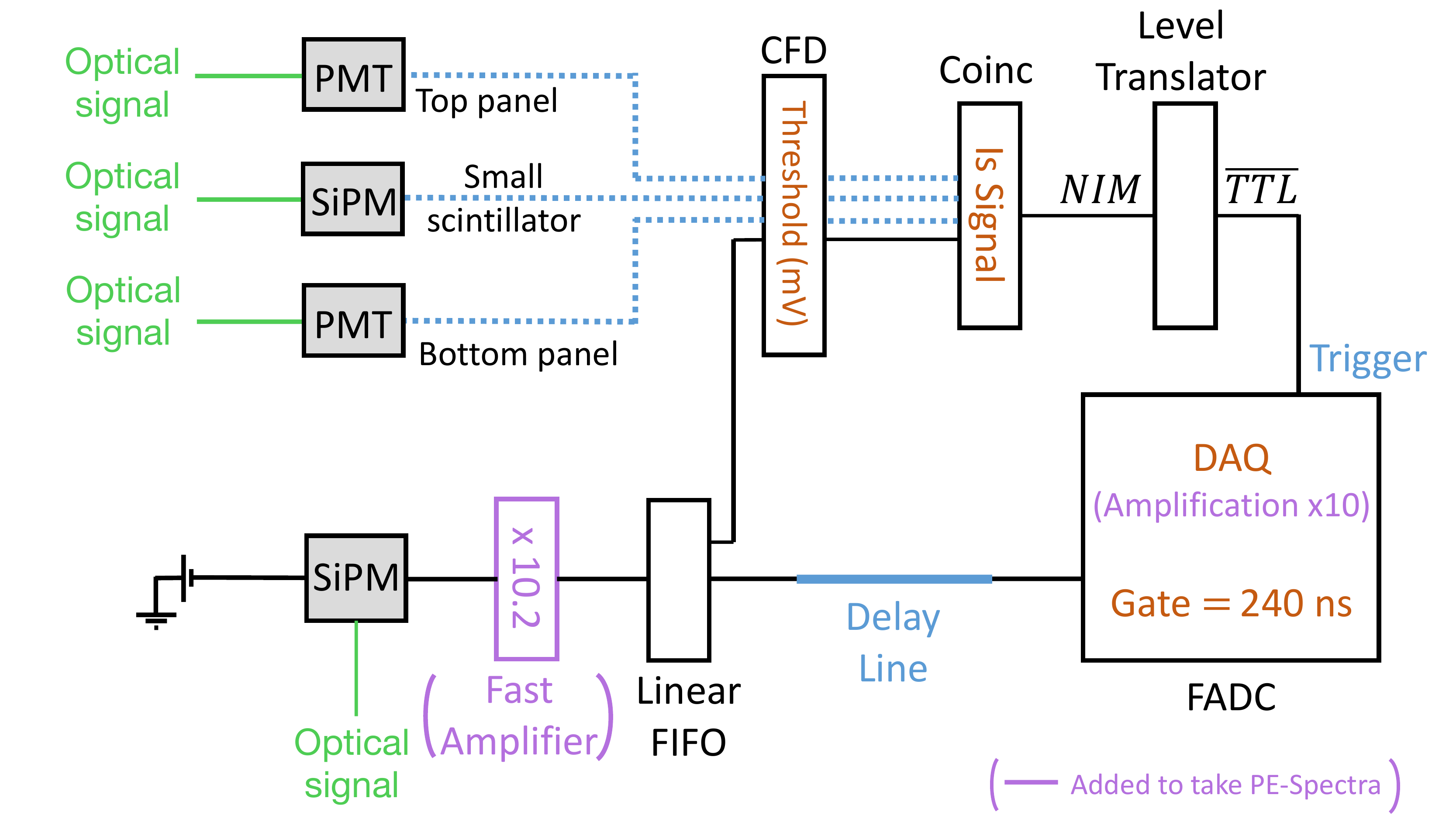}
    \caption{DAQ scheme. The prototype SiPM signals are processed by an FADC. The external trigger is generated by a constant fraction discriminator (CFD), a coincidence module (Coinc) and a level translator. By demanding a coincidence of the prototype SiPM(s) with the top and bottom PMT, or the SiPM of the small plastic scintillator, different muon samples can be selected. Optionally, the SiPM signals can be amplified by a fast amplifier and an FADC-internal amplification used in gain calibration measurements.}
    \label{fig:DAQ_scheme}
\end{figure}

   Figure~\ref{fig:DAQ_scheme} shows a scheme of the data acquisition (DAQ). 
   The signals are split by a linear fan-in fan-out (FIFO), of which one output signal is used for trigger generation and the second one is fed into a 2-channel fast analog-to-digital converter (FADC) with a sampling rate of 250\,MHz. 
   Optionally, the signals can be amplified with a fast amplifier providing an amplification factor of 10.2, and an internal gain of 10 provided by the FADC card. 
   For simplicity only the integrated charge of the SiPM pulse within a 200\,ns window following an external trigger is recorded. The charge is corrected online for the integrated baseline, which is calculated in a 40\,ns window prior to the trigger. The trigger is generated by analog NIM electronics, composed of a constant fraction discriminator (CFD), a coincidence module and a level translator. For the straight fiber configuration, a coincidence of the two SiPMs is required. 
   Depending on the coincidence condition different data sets are recorded: a sample of atmospheric muons is selected by a three-fold coincidence of the prototype, the top and the bottom panel. Muons passing through a restricted volume of the prototype are selected by a coincidence of the prototype, the bottom panel and the small plastic scintillator.  
   
	\section{Calibration of the SiPMs}
	\label{sec:Calib}

	\begin{figure}[h!]
    \centering
    \noindent\makebox[\textwidth]{
    \begin{subfigure}[b]{0.49\textwidth}
        \includegraphics[width=\textwidth]{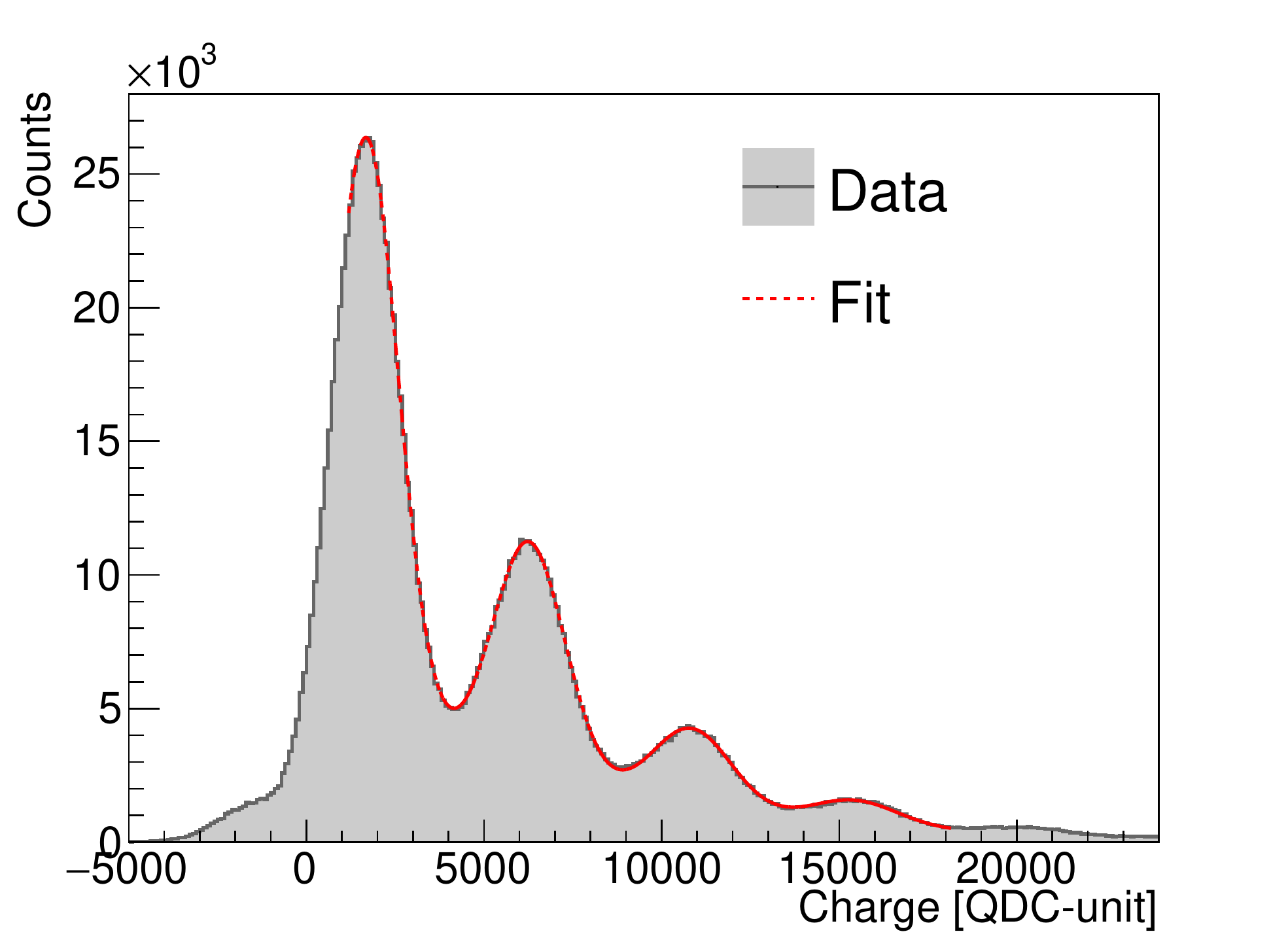}
    \caption{\small{PE spectrum}}
    \label{fig:fitted_spectra}
    \end{subfigure}
    \begin{subfigure}[b]{0.49\textwidth}
        \includegraphics[width=\textwidth]{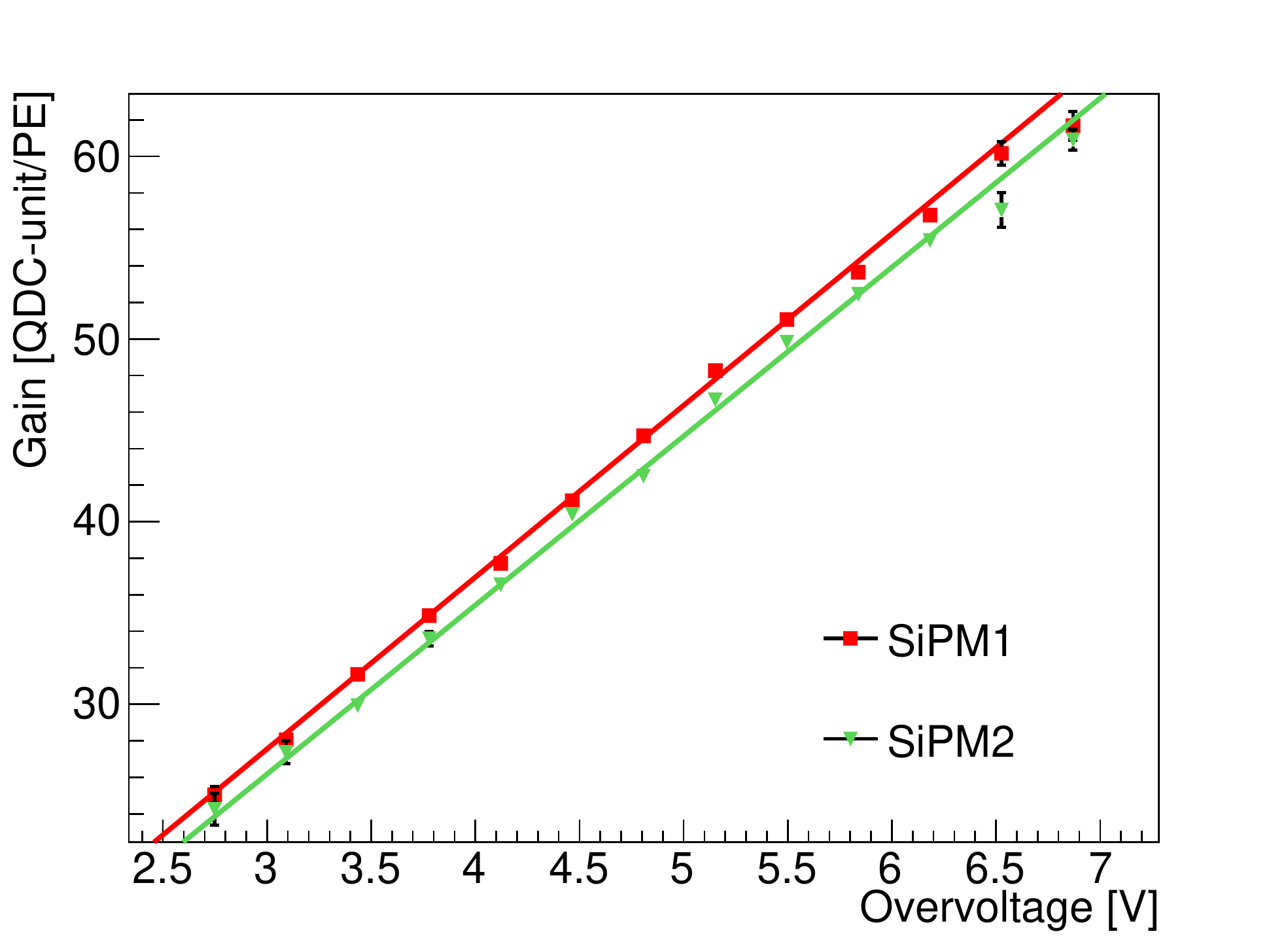}
    \caption{\small{PE calibration curves}}
    \label{fig:calib_curves}
    \end{subfigure}
    }
    \caption{(\subref{fig:fitted_spectra}) shows the charge distribution of the photoelectron (PE) spectrum measured at an overvoltage of 4.8\,V. The data are taken with an additional gain of 102, provided by the FADC card and the fast amplifier, see figure~\ref{fig:DAQ_scheme}. The charge spectrum is fitted with a general fitting model to extract the gain, i.e. charge (in QDC units) per PE. (\subref{fig:calib_curves}) shows the gain calibration curves of the two SiPMs as a function of the overvoltage. For the characterization measurements, a working point of 4.8\,V was chosen. Since the characterization data presented in section~\ref{sec:performance} are taken without the additional amplification, the gain curve is corrected for it.}
\end{figure}
	
	SiPMs are arrays of independent avalanche photo diodes (APD) operated in Geiger mode. 
A photon which is absorbed in an APD can trigger an ionization cascade. When applying a strong enough electric field inside the silicon, a so-called Geiger discharge is initiated, and the initial electron-hole pair produced in the interaction of the photon is converted into a macroscopic current flow. The bias voltage above which this process becomes possible is the breakdown voltage, and lies between 24\,V and 25\,V for the PE3325-WB-TIA-SP SiPM modules operated at 21$^{\circ}$C~\cite{KETEK}.
The gain of a SiPM gives the amount of charge created by each detected photon and is a function of the applied overvoltage, i.e. the difference between applied bias voltage and the breakdown voltage. 
	
For the gain determination, we made use of the large dark count rate of the SiPMs. At an overvoltage of 5\,V we expect a rate of $\mathcal{O}(1\,\text{MHz})$ thermally initiated Geiger discharges at $21^{\circ}$C~\cite{KETEK}. 
The signals of Geiger discharges originating from thermally generated electrons are identical to the ones generated by photon absorption.   
In order to record spectra with single or few Geiger discharges, the SiPM signals are amplified by combining the fast amplifier and the internal amplification of the FADC shown in figure~\ref{fig:DAQ_scheme}. A low CFD threshold value of 1\,mV is chosen in order to trigger the DAQ on low-charge events from the dark count rate. Figure~\ref{fig:fitted_spectra} shows a charge spectrum recorded from dark counts at an overvoltage of 4.8\,V.
Peaks corresponding to 1, 2, 3 and 4 Geiger discharges are clearly seen. The gain is given by the distance between two consecutive peaks. 
To extract the gain, we fit the charge spectrum with a slightly modified model presented in~\cite{Chmill:2016ghf} where we add a general exponentially decreasing background. The model takes into account the electronic noise of the system, the number of initiated Geiger charges, prompt and delayed cross-talk among the pixels, and after pulses. 
Figure~\ref{fig:calib_curves} shows the extracted gain for different bias voltages for the two SiPMs used with the prototype. As expected, the gain increases linearly with the overvoltage. 
The reconstructed linear fit corresponds to the gain calibration curve, which is used to convert the measured charge of the integrated SiPM signal into the number of detected photons. In analogy to PMTs the term number of photoelectrons (PE) will be used in the following. 

The calibration of the measured spectra has two advantages: it allows for a direct comparison of the different fiber configurations using independent SiPMs, and a monitoring of the stability of the SiPMs' performance. 
The latter is a key requirement for the long-term operation of the \NUCLEUS{}muon veto with a constant high efficiency.  
We observe no difference in the gain between measurements in which the WLS fibers are attached to the SiPMs compared to the fibers detached. Therefore, we conclude that the stability of the final \NUCLEUS{}muon veto modules can be monitored by regular 1-minute gain calibration measurements with WLS fibers attached. Neither hardware changes of the modules themselves nor the installation of an LED-based calibration system inside the light tight boxes are needed.

	
	\section{Performance of different fiber configurations of a single scintillator module}
	\label{sec:performance}
	
		\begin{figure}[t!]
    \centering
    \noindent\makebox[\textwidth]{
    \begin{subfigure}[b]{0.49\textwidth}
        \includegraphics[width=\textwidth]{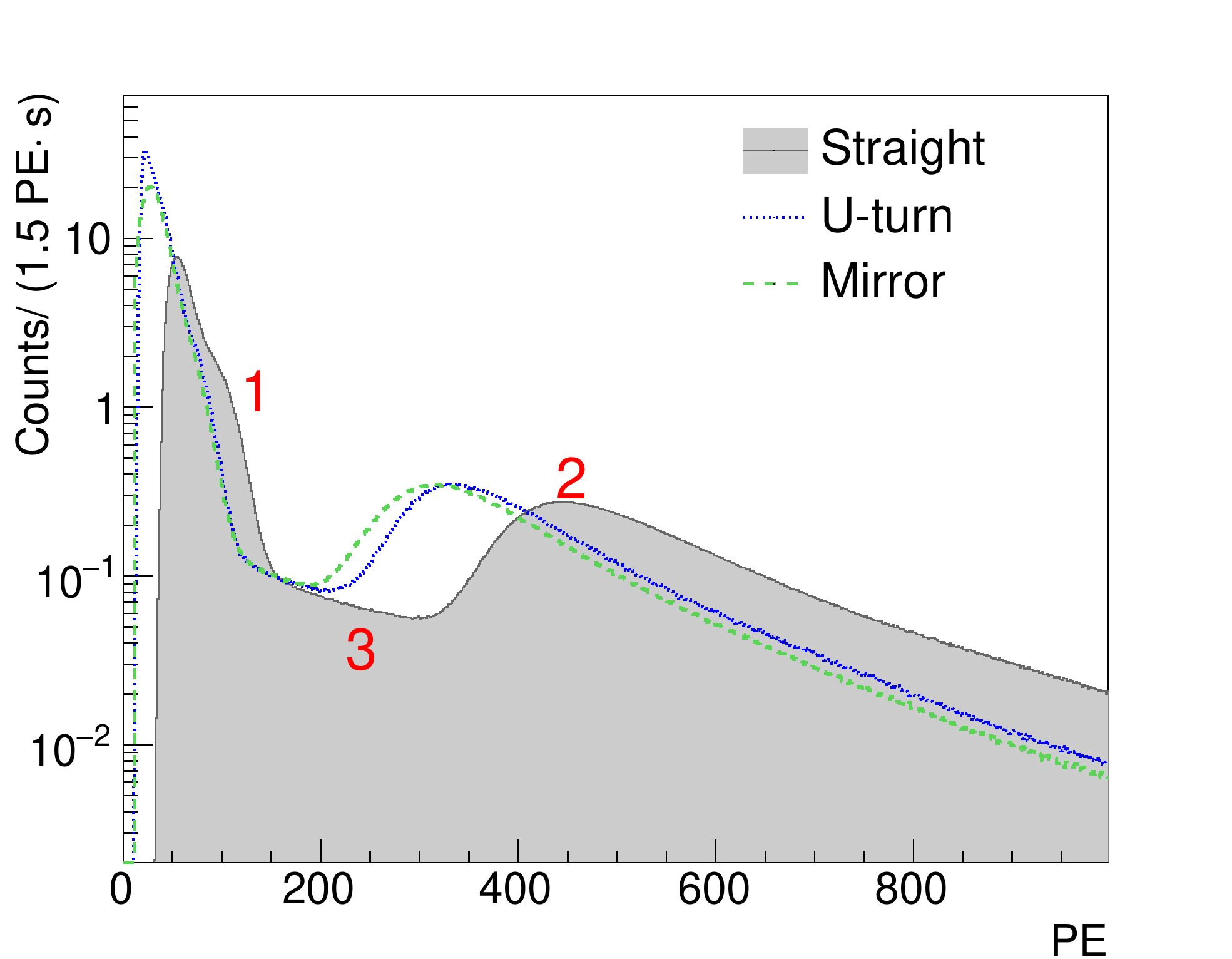}
    \caption{\small{Full spectra of different configurations}}
    \label{fig:muonBck_spectra}
    \end{subfigure}
    \begin{subfigure}[b]{0.49\textwidth}
        \includegraphics[width=\textwidth]{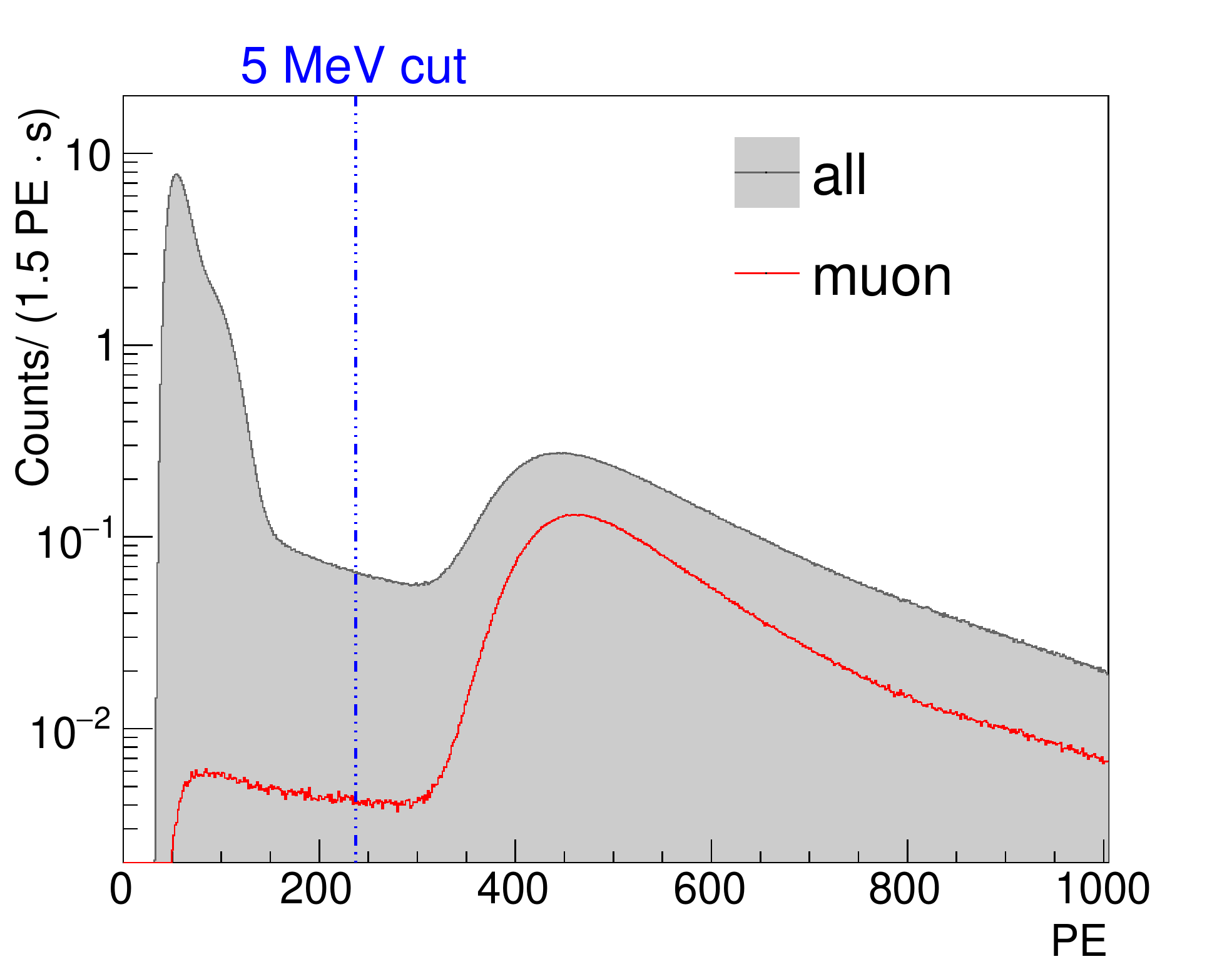}
    \caption{\small{Muon spectrum of the straight fiber configuration}}
    \label{fig:muon_spectra}
    \end{subfigure}
    }
   \caption{(\subref{fig:muonBck_spectra}) shows the full spectra measured with the different fiber configurations. The threshold is set to 10\,mV. For the straight fiber configuration a coincidence in the two SiPMs is required. All spectra show the same features: (1) an external $\gamma$-background, (2) Landau-like distributed muon events and (3) a well defined "plateau" separating these two contributions. In~(\subref{fig:muon_spectra}) the full spectrum (gray area) is compared to the muon spectrum (red line) both measured in the straight fiber configuration. Muon events are identified by means of a three-fold coincidence in the \NUCLEUS{}prototype, the top and the bottom panel of the test stand. The muon identification power is defined as the fraction of selected muon events above a certain threshold, here 5\,MeV.}
    \label{fig:Spectra}
\end{figure}

	We fully characterized the three different fiber configurations illustrated in figure~\ref{fig:configurations}. The overall light yield and the  separation  of the ambient $\gamma$-background from muon events are of central interest for the performance of the muon veto panels. Therefore, as a first step we characterized the full spectra of events in the prototype panel triggering at a low threshold of 10\,mV.
	Figure~\ref{fig:muonBck_spectra} shows the measured spectra. For the straight fiber configuration the sum of the calibrated SiPMs is used. 
	All three spectra show the same features: (1) an external $\gamma$-background sharply increasing towards the threshold, (2) muon events following a Landau-like distribution with the most probable value corresponding to vertical muons passing through the 5\,cm thick plastic scintillator and (3) a well defined "plateau" separating these two contributions.
	As expected, the straight fiber configuration yields the highest light output. The spectra of the U-turn and the mirror configuration are shifted to lower PE values. However, the photo-statistics for muon events remains large as well as their separation from the $\gamma$-background.
	
	In the following, we quantify the light yield, the separation of  muon events from ambient $\gamma$-background, the muon identification power and the homogeneity of light collection of the prototype panel in different fiber configurations. 
	The purpose of this study is to understand the differences between the fiber configurations in view of their use in a compact and efficient assembly of panels covering the full solid angle around the central target detectors of the \NUCLEUS{} experiment.

	\begin{table}[t!]
	   \centering
	    \caption{Summary of the performance of the different fiber configurations. The total light yield (LY) and the peak to valley ratio (P/V) are shown. Only statistical uncertainties are given. $\Delta\epsilon$ gives the relative muon identification power compared to the straight fiber configuration. }
	   \begin{tabular}{l|c|c|c}
	   \toprule
	        & LY & P/V & $\Delta\epsilon$ \\
	       Configuration & [NPE/MeV] & & \\
	       \midrule
	        Straight fiber & 47.47$\pm$0.02 & 4.86$\pm$0.06 & 1.00 \\ 
	        U-turn  & 32.13$\pm$0.02 & 4.25$\pm$0.04 & 0.99 \\
	        Straight fibers + mirror & 30.24$\pm$0.02 & 3.89$\pm$0.03 & 0.99 \\
	         \bottomrule
	    \end{tabular}
	    \label{tab:performance}
	\end{table}

	\subsection{Light yield}
	\label{sec:LY}

We define the light yield (LY) as the number of recorded Geiger discharges (or NPE) per MeV. Table~\ref{tab:performance} summarizes the LY for the different fiber configurations which is determined by the position of the measured muon peak in NPE over the expected position in MeV. The latter is determined by a dedicated GEANT4 Monte Carlo simulation~\cite{GEANT4:2002zbu} of the full test setup.   
	As expected, the straight fiber configuration shows the highest LY with 47.5~PE/MeV. The LY is reduced by 32~\% (36~\%) for the U-turn (mirror) configuration. 
	
	The separation of the $\gamma$-background and the muon distribution can be quantified with the peak-to-valley ratio (P/V), the ratio of the rate at the maximum of the muon peak over the rate at the minimum of the plateau. 
	The rates are determined by fitting the plateau region with an 8th degree polynomial and the muon peak region with a Landau distribution. Table~\ref{tab:performance} summarizes the results for the three different fiber configurations.

	\subsection{Muon identification power}
	\label{sec:Efficiency}

To select a distinct sample of atmospheric muon events, we record only events which are in coincidence with the top and bottom panel, see figure~\ref{fig:DAQ_scheme}. Figure~\ref{fig:muon_spectra} shows the spectrum of selected muon events (red line) compared to the full spectrum (gray area) measured in the straight fiber configuration. 
	The condition of coincidence with the top and bottom panels restricts the angular acceptance of the muons and hence their total rate. The suppression of the $\gamma$-background reveals the full structure of the plateau. 
	This plateau is dominated by muon events traversing the prototype close to an edge. Such events feature a reduced track length through the plastic scintillator and hence deposit less energy. Furthermore, the light collection of edge events might be reduced with respect to muons passing the core of the plastic scintillator. 
	
The muon identification power, $\epsilon$, is defined as the fraction of selected muon events above a certain threshold. 
The straight fiber configuration exhibits an identification power of $\epsilon=(97.26\pm0.01)$\% for a cut set at an equivalent of 5\,MeV, see figure~\ref{fig:muon_spectra}.
The chosen cut is well above the highest natural $\gamma$-line of $^{208}$Tl at 2.6\,MeV. 
While the LY and P/V are significantly increased for the straight fiber configuration, the muon identification power differs only by $\sim$1\% among the different configurations, see table\ref{tab:performance}.
The study validates that the mirror and U-turn configuration reach a similar performance as the straight fiber configuration. 

The above study highly depends on the selected muon sample and, thus, on the geometry and the imposed trigger conditions. Hence, the muon identification power of a single panel cannot be scaled to the overall performance of the final \NUCLEUS{}muon veto. With the three-fold coincidence, we select a large number of \textit{plateau}-events which clip an edge of the panel. Missing to identify such events in a single panel may not be relevant in the final muon veto configuration since (a) such muons may not pass the passive shielding or (b) may be detected in a neighboring panel. 
In the final \NUCLEUS{}muon veto, the panels will be packed very closely, see section~\ref{sec:SummaryConfigs}. Thus, muons clipping the edge of the plastic scintillator, may leave a longer track length in the neighboring panel, and hence deposit more energy. The geometrical efficiency will be given in section~\ref{sec:GeoEff}, the total muon identification efficiency will be determined during the commissioning of the \NUCLEUS{}muon veto in 2022.

	\subsection{Light yield homogeneity}
	\label{sec:LYHomo}
	
	\begin{figure}[!b]
\centering
    \includegraphics[width=0.49\textwidth]{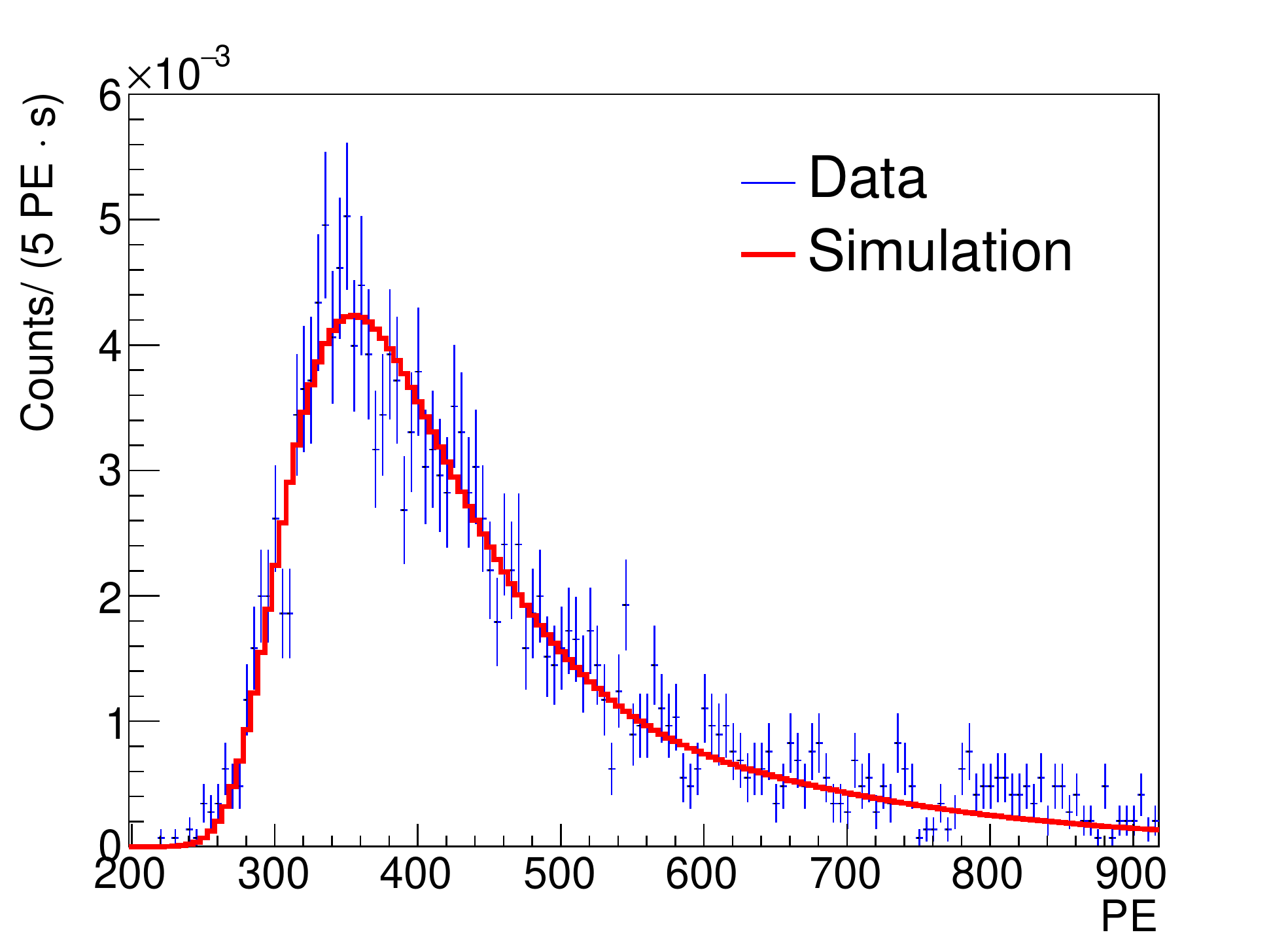}
    \caption{Comparison of local muon charge spectra from the measurement at the central position in the U-turn configuration and the folded simulation. The fitted conversion factor, or local light yield, is $\eta=(33.3\pm0.2)$\,PE/MeV, for a $\chi^2/\text{dof}=97.5/87$.}
    \label{fig:Muon_simulation}
\end{figure}
	
	Sizeable differences in the LY throughout the muon veto module may result in local differences in the muon identification power. Therefore, in addition to the total LY, presented in section \ref{sec:LY}, we measured the local LY for the different fiber configurations. The aim is to identify potential sources of LY inhomogeneity attributed to the fiber configurations.  

The LY is defined as the number of PE detected in the SiPM over the average deposited energy, and can be influenced by local differences in the collection of the scintillation light by the fibers and by losses when the light is subsequently propagated to the SiPM. 
The former may e.g. be affected by a local variation of the optical coupling between fibers and plastic scintillator, or by the distance to a side face or to the pool. The latter may occur due to bending of the fibers, or poor optical couplings of the fiber to the mirror (if used) or to the SiPM. Furthermore, as the trapped light is attenuated, the fiber length influences the fraction of light reaching the SiPM.

To measure the local LY, we select a sample of muons by means of a three-fold coincidence in the prototype (middle panel), the bottom panel and the small plastic scintillator shown in figure~\ref{fig:Setup_scheme}. Setting the position of the small scintillator following the grid displayed on figure~\ref{fig:double_sided} and \ref{fig:U-turn}, we can directly measure the PE spectrum from atmospheric muons depositing their energy within a local volume of the prototype. From muon track simulations analogous to the ones described in section \ref{sec:GeoEff}, we assess that 90\% of the atmospheric muons interacting through this 3-fold coincidence scheme deposit their energy in the prototype within a truncated conic volume of $\diameter \ 12$ cm $\times \ \diameter \ 22$ cm $\times \ 5$ cm, see figure~\ref{fig:Setup_scheme}.

As we expect a position dependence of the average energy deposition by the selected muons, we performed a GEANT4 Monte Carlo simulation~\cite{GEANT4:2002zbu} of atmospheric muons interacting in the setup including the three-fold coincidence. Such differences originate in the different thickness of the plastic scintillator in the pool region, as well as shorter track lengths of the crossing muons close to the edges with respect to the central position. To extract the interesting physical quantity of the local light yield, we use these predicted energy depositions as a basis of a simple model to fit the measured local muon PE spectrum through the least squares method. This model has only two free parameters: a global normalization factor and an $\eta$ parameter used as the mean and the variance of a Gaussian response function, thus accounting for the local light yield and its associated Poissonian fluctuations. Figure~\ref{fig:Muon_simulation} demonstrates that we obtain a good agreement between data and simulation. 

\begin{figure}[!t]
\centering
    \includegraphics[width=0.9\textwidth]{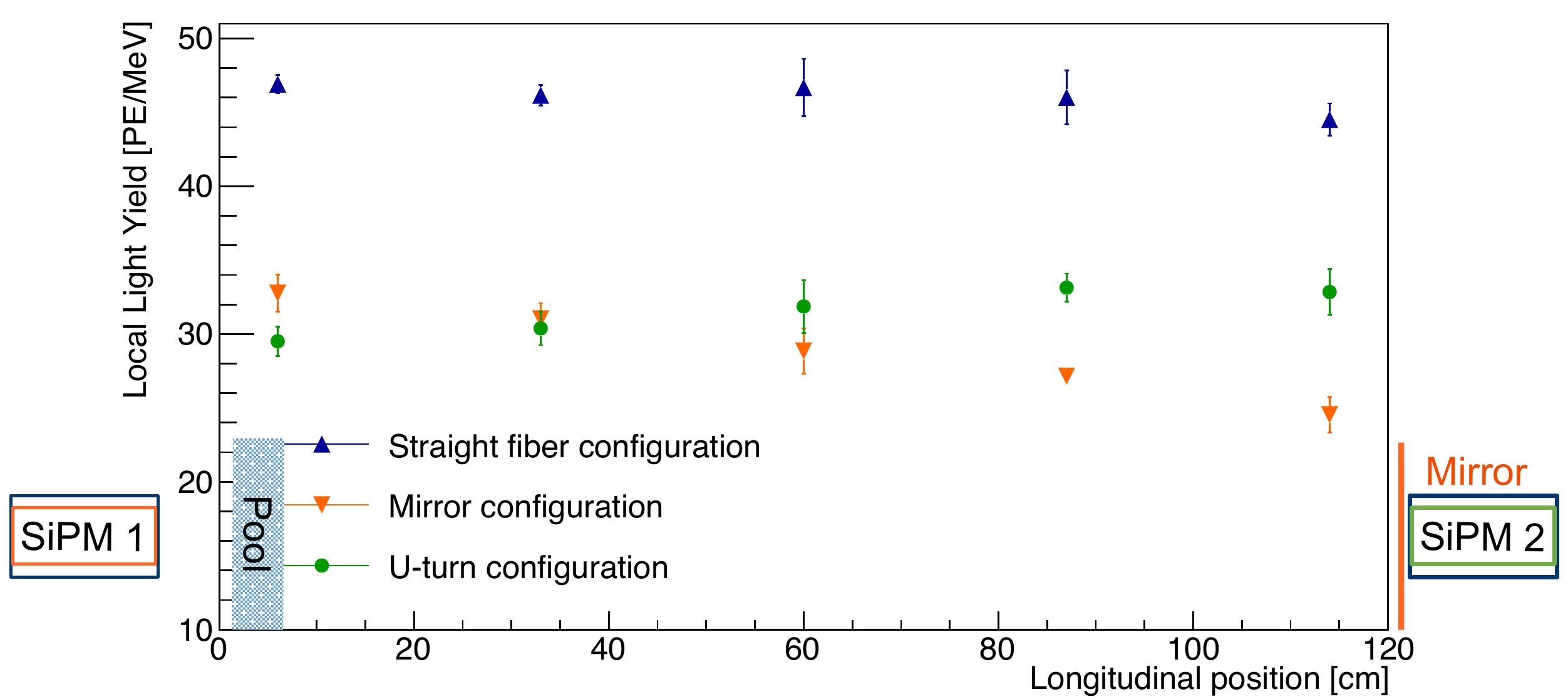}
    \caption{Local Light Yield in the prototype read out with different fiber configurations. We average over the three transverse positions at same longitudinal position. The uncertainties include the spread of the measurements at a given longitudinal position. For the measurement in the straight fiber configuration SiPM1 and SiPM2 are used, for the mirror configuration, SiPM2 is replaced by a mirror, and the U-turn configuration was performed with SiPM2 respectively.}  
    \label{fig:LYH_Mu}
\end{figure}

In figure~\ref{fig:LYH_Mu}, we report the results for the extracted local light yield of the different fiber configurations. For simplicity we average over three transverse positions at the same longitudinal position. 
The straight fiber configuration shows the least inhomogeneities in the LY: the longitudinal positions from 6\,cm to 87\,cm show a 2\% variation. The drop in LY at 114\,cm, close to SiPM2, may hint towards additional light losses at the side of  SiPM2, e.g. by poor coupling of the WLS fibers to the SiPM. 
The mirror configuration shows a 30\%-level inhomogeneity in the LY. As expected, the LY increases towards the SiPM used for read-out, and worsens towards the mirror. This effect can be largely attributed to light attenuation in the fibers. Similarly, the U-turn configuration shows an 11\%-level inhomogeneity along the prototype, with the LY being reduced towards the U-turn, i.e. the pool region. This result hints towards a significant reduction of losses during light propagation in the U-turn fiber configuration with respect to the use of a mirror, regardless of the additional bending of the fibers.

	\subsection{Selection of fiber configuration for the final \NUCLEUS{}muon veto}
	\label{sec:SummaryConfigs}
	
	For a successful operation of \textsc{Nucleus}, a 4$\pi$ coverage of the passive shielding by the muon veto is essential. Each of the muon veto modules carries the plastic scintillator (active volume) and up to two SiPMs, depending on the fiber configuration. Due to the minimal bending radius of 5\,cm of the optical fibers, a distance of about 17\,cm between the scintillator and the SiPM is required. Thus, each SiPM introduces a large non-instrumented volume inside the module. 
	
\begin{figure}[t!]
        \centering
        \includegraphics[width=13cm]{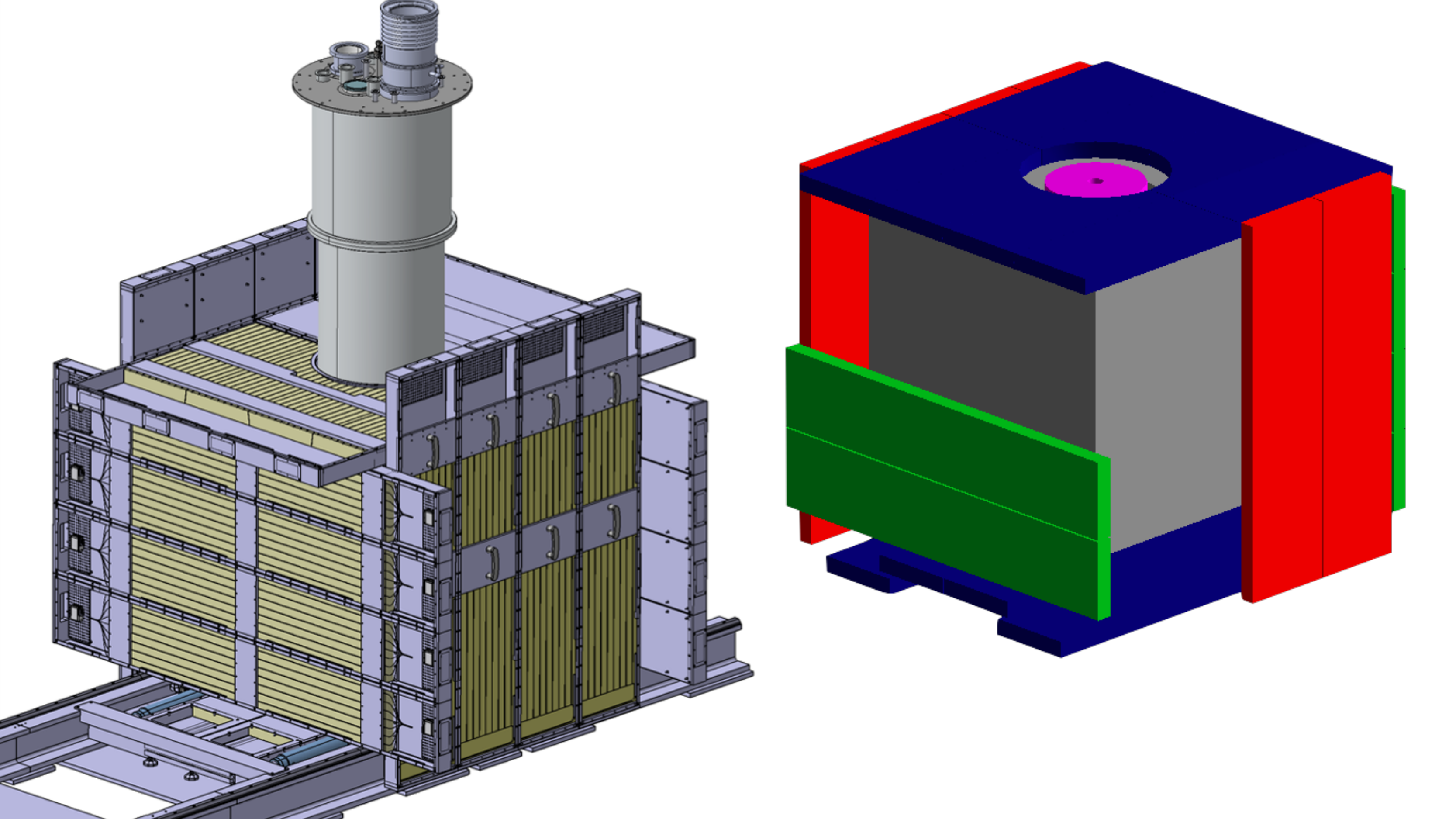} 
        \caption{On the left, a 3D view of the arrangement of the 28 scintillator panels (displayed in yellow color) forming the hermetic muon veto of the \NUCLEUS{}experiment is shown. For an optimal detection of muons, the bottom part and two of the four side walls are readout by the two-sided straight configuration, the shorter top panels  with a one-sided straight configuration, and the remaining two side walls are equipped with a U-turn fiber configuration. The shielding is placed on rails in order to access the target detectors placed inside the cryostat. On the right, the geometrical arrangement as implemented in the track simulation is shown. The inner active shielding (i.e. the cryogenic muon veto), shown in pink, is positioned on the same level as the top panel. The passive lead shielding is shown in grey.}
        \label{fig:3D_View}
    \end{figure}

	The presented studies show that all three fiber configurations have similar performance. 
   The major challenge for the final design of the \NUCLEUS{}muon veto is the optimal coverage of the passive shielding with (active) plastic scintillator to avoid coverage gaps. 
   The left part of figure~\ref{fig:3D_View} shows the default arrangement of the individual muon veto modules hermetically arranged around the \NUCLEUS{}cryostat and the passive shielding:
    \begin{itemize}
        \item We pay special attention to the top part of the veto. Since muons pass through the experimental setup mainly from the top, any coverage gap in this part needs to be avoided. At the same time, we need to free space for the cryostat itself and  ensure the access to it. In order to be able to open the shielding, the top part consists of two large stainless steel boxes. Each box contains four panels with trapezoidal cross-section. This way, the spacing of the neighboring plastic panels is minimized to the thickness of the diffusive foil of 200~$\mu$m and not aligned with the vertical direction of most muons. 
        The top muon veto part features a hole with a diameter of 456\,mm where the cryostat containing the \NUCLEUS{}detectors enters the experimental setup. The impact of this hole shall be mitigated by operating a disk-shaped active muon veto inside the cryostat at sub-Kelvin temperatures equally instrumented with WLS fibers and a SiPM~\cite{MT:Erhart2021}. 
        Only a 6.5~cm-wide annular gap remains between the muon veto boxes and the inner most vessel of the cryostat. Muons entering through this gap will be efficiently vetoed by the side and bottom modules. Each top panel is read out with a simple straight fiber configuration. As the panels are only 62.8~cm long at most, we expect the light yield inhomogeneity due to attenuation effects to be much reduced. For the moment, we keep the option to place a mirror at the open end of the fibers to recover part of the scintillation light. 
        \item The front/rear (left/right) side walls consist of 8 (8) modules, each featuring a $24.3\times5.0\times125.4$\,cm$^3$ ($28.2\times5.0\times118$\,cm$^3$) plastic scintillator hosted inside a separate stainless steel box.
        The usage of individual boxes eases mounting. Still it is not possible to mount all modules with the same orientation without introducing large gaps. Therefore, the modules of two sides are oriented horizontally, while they are oriented vertically for the other two sides. The former feature a two-sided straight fiber configuration, while the vertical modules are in U-turn configuration with the SiPM at its top part. In such a configuration the 4-5 mm gap between neighboring active pieces of plastic scintillator inside their respective box intercepts a fully negligible phase space of the muon flux. This same configuration also minimizes gaps at the bottom of the muon veto side increasing the efficiency to detect muons which enter the experimental setup through the unavoidable gap between cryostat and shielding. 
        \item The bottom part consists of four identical  $25.9\times5.0\times124.3$\,cm$^3$ plastic scintillator panels hosted inside a separate stainless steel box read out at both ends with the straight fiber configuration. The modules are placed inside the custom-made rail system the shielding is placed on. The bottom panels are essential to veto muons entering the shielding through the gaps in the top muon veto.  
    \end{itemize}

	\section{Determination of the geometrical efficiency}
	\label{sec:GeoEff}
	
	A Monte Carlo simulation tool implemented in ROOT~\cite{ROOT} was used to estimate the geometrical efficiency and the muon count rate of the full assembly of the \NUCLEUS{}muon veto. The TGeometry package~\cite{TGeo} was used for building, navigating and visualizing the detector geometries. The performed study was of purely geometrical nature in the sense that the generated muons travel in straight tracks throughout the complete simulation process without taking physical interactions into account. 
	Still, these track simulations provide a suitable tool to assess the coverage of the passive shielding and draw attention to gaps in the overall arrangement that lead to a loss of efficiency. 
		
\begin{figure}[h!]
       \centering
      \noindent\makebox[\textwidth]{
      \includegraphics[width=0.75\textwidth]{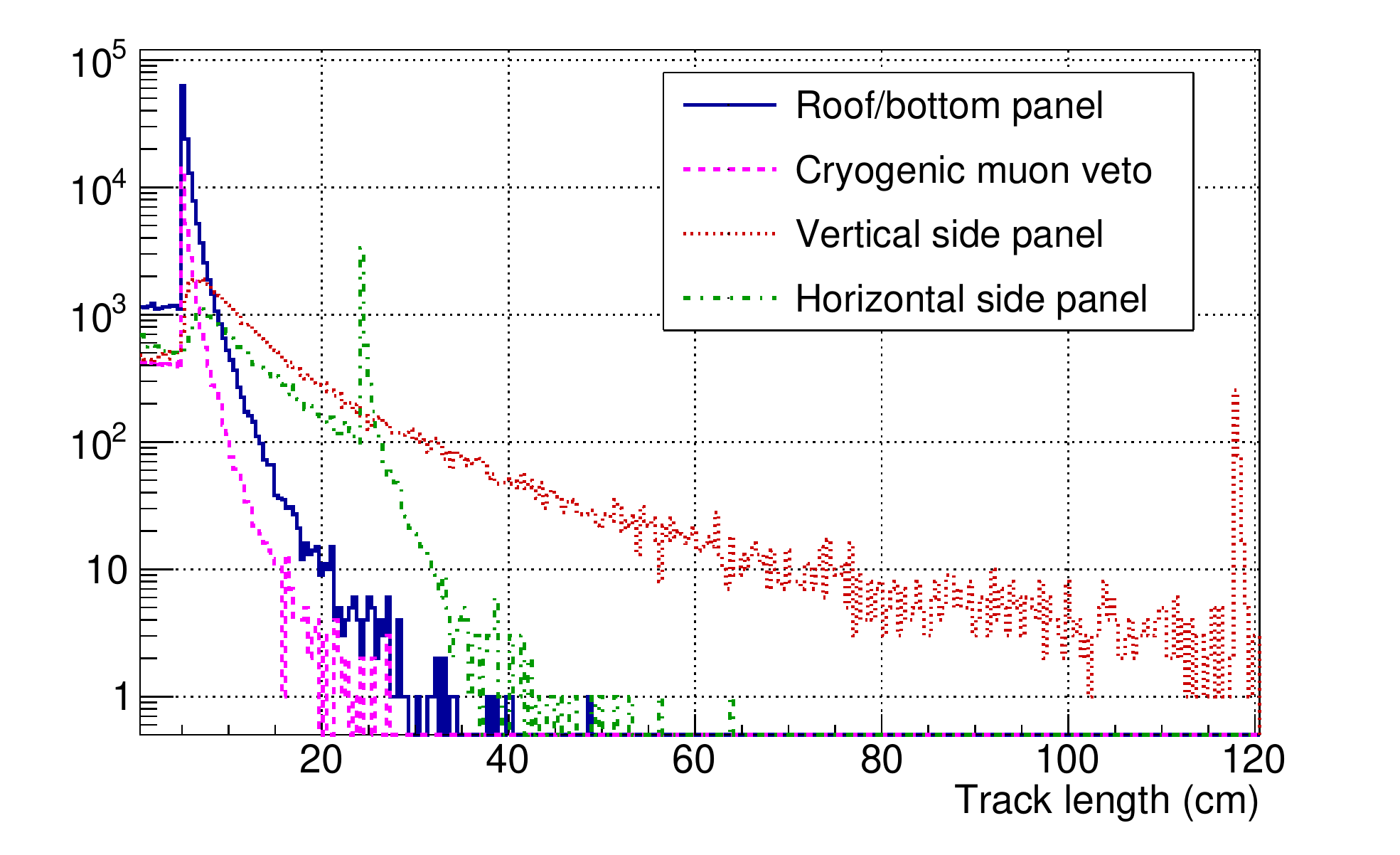}
        }
      \caption{Track length spectra for the different panels composing the muon veto. The color code of the panels corresponds to the one shown in figure~\ref{fig:3D_View}.}
     \label{fig:track_spectra_cm}
    \end{figure}

	The implemented geometry was modeled according to latest technical drawings of the individual \NUCLEUS{}components (see figure~\ref{fig:3D_View} right). The configuration is explained in detail in section~\ref{sec:SummaryConfigs}.  
The track simulations provide a simple, but powerful tool to assess the overall geometrical coverage of the muon veto and estimate the expected muon rate.  
Figure~\ref{fig:track_spectra_cm} shows the prediction of the track-length distributions for the different orientations of the scintillator panels. For all horizontal panels (roof, bottom plane and cryogenic muon veto disk) we obtain distributions with a prominent peak at 5\,cm track-length corresponding to the thickness of the scintillator. In the side wall panels the distributions evolve towards longer track-lengths. The main axis of these panels being either horizontal or vertical, the position of the peak value shifts to the width value (24.1\,cm) or the total length value (116.0\,cm) respectively. 
All configurations show a plateau between 0\,cm and 5\,cm, as measured with the prototype with intensities well below the muon peak. 
We expect a clear separation between $\gamma$-rays and muons events as previously observed with the prototype, and that a similar threshold of 5\,MeV, corresponding to 2.5\,cm, can be reasonably applied in the following. 

From the track-length distributions we can define the muon geometrical efficiency as the ratio of detected muons to the total number of muons entering the experimental setup. A muon is considered as "detected" if its track-length in at least one panel is greater than 2.5\,cm. For a muon at the minimum ionisation energy loss this value corresponds to the 5\,MeV threshold discussed above. The notion of "entering the experimental setup" requires to define a volume inside which any muon will be considered as a potential source of background. In the case of the \NUCLEUS{}experiment, the natural volume is the lead shielding, which is surrounded by the scintillator panels. This volume defines the proximity of the cryogenic detectors at the center of the setup and at the same time contains all the dense material of the apparatus in which muons can generate secondary neutrons by spallation reactions. The cube assembly of all panels being quite hermetic, it is very unlikely that a muon enters the lead shielding without traversing at least 2.5\,cm in one of the scintillators. We obtain a geometrical  efficiency of 99.7\%, meeting the specification of the \NUCLEUS{}experiment. The efficiency loss can be attributed to muons which clip the plastic scintillators at an edge leaving a track length of less than 2.5\,cm, as well as  muons traversing the setup through the remaining small gaps in the \NUCLEUS{}muon veto shielding without crossing any panel. 
    
    The track simulations can also be used to assess the acceptable  uncertainty on the trigger threshold, as e.g. induced by variation in the LY along the panel. In section~\ref{sec:LYHomo} we measured an 11\% inhomogeneity in the LY with the U-turn fiber configuration. A reduced local light yield translates into a higher effective threshold. Assuming a 15\% variation in the LY, and correspondingly increasing the threshold from 2.5\,cm to 2.875\,cm, changes the muon efficiency by $<$0.1\%. Thus, we conclude that even for large variations of $\mathcal{O}$(10-15\%) in the LY with respect to the central position, the effect on the muon efficiency is at the sub-0.1\% level.

  The total muon rate expected in this detector is determined by normalizing our simulation to the expected muon flux at the \NUCLEUS{}experimental site, the VNS: we correct the integral flux of vertical muons above 1\,GeV at sea level of $\Phi=70\,\mu/(\text{sr}\cdot\text{s}\cdot\text{m}^2)$~\cite{Patrignani:2016} for the angular dependence of the intensity and extrapolate the  integral spectrum down to 0.1\,GeV. The VNS provides a muon attenuation factor of 1.41~\cite{Angloher:2019flc}. Thus, for the \NUCLEUS{}muon veto placed at the VNS,  we predict a total rate of 325 identified muons/s. 
    Assuming a 50\,$\mu$s veto time-window centered on the arrival time of each identified muon~\cite{Angloher:2019flc}, the induced dead-time of 1.6\% remains moderate, meeting the specification of the experiment.

	\section{Conclusion and outlook} 
	\label{sec:conclusion}
	
	Muon-induced events are expected to be one of the dominant background sources for the \NUCLEUS{}experiment. 
	For a successful \cevns{} measurement, it is of great importance to efficiently mitigate this background by means of a  muon veto hermetically surrounding the experimental setup.  
	Therefore, we developed compact and efficient muon veto modules consisting of  5\,cm thick BC-408 plastic scintillator panels equipped with WLS fibers and wrapped in diffusive foil. The flexibility of the fibers allows for different read-out configurations. In a dedicated test stand, we fully characterized three different fiber layouts to collect and guide the scintillation light towards SiPMs, see figure~\ref{fig:configurations}: 
	(1) The baseline design features nine WLS fibers that run in straight grooves along the full length of the panel, and two SiPMs mounted at each front side of the scintillator (straight configuration). 
	(2) In the mirror configuration, one SiPM is replaced by a mirror. 
	(3) In the last arrangement, the fibers describe a U-turn at one front side of the panel and the light is collected with a single SiPM (U-turn configuration).

	All three configurations feature a light yield of $>$30\,PE/MeV and high separation power of muon- from $\gamma$-events. 
    The straight fiber configuration features a 2\% inhomogeneity of the LY over the full panel, while the U-turn configuration shows an 11\% variation of the detector response, probably dominated by attenuation effects in the fibers. 
    The reduction of the read-out to a single SiPM reduces costs related to typically expensive read-out channels. Furthermore, the U-turn configuration has the additional advantage of an even more compact design than the straight configuration at a similar muon identification power. We disregard the mirror configuration due to the large observed 30\% inhomogeneity of the LY, which we attribute mostly to attenuation effects in the fiber and poor optical coupling of the mirror. 
	
	In figure~\ref{fig:3D_View} we present the design of the \NUCLEUS{}muon veto. We foresee an arrangement of individual modules with different fiber configurations tightly packed around the passive shielding: for the bottom, the top and two of the four side walls we will use the straight fiber configuration, and for the remaining side walls the U-turn configuration respectively. With this design, we maximize the coverage of the passive shielding by the active muon veto. 
	The \NUCLEUS{}muon veto is completed with a disk-shaped cryogenic muon veto hosted inside the cryostat. 
	Based on geometrical simulations, we estimate the geometrical efficiency of the muon veto to  99.7\%. 
	The estimated muon rate is 325\,Hz, inducing a dead time of 1.6\% in the cryogenic target detectors. 
	
	The test stand which was constructed for this work will be used to test the individual panels prior to their installation in the \NUCLEUS{}setup and, hence, ensure a nominal performance.  
	The overall muon identification efficiency of the muon veto will be determined during the \NUCLEUS{}commissioning phase, which is planned to start in summer 2022 in the shallow underground laboratory UGL at the Technical University of Munich.


	\acknowledgments
	We are grateful to the technical staff of our laboratories for their excellent work in designing the \NUCLEUS{}muon veto and for their contribution to this work. 
	We thank Nicole D'Hose and Etienne Burtin from the COMPASS collaboration, as well as Patrick Champion, Olivier Leseigneur and Yann Reinert from CEA-Saclay for the great support during the R\&D phase and their support to recover the plastic scintillators. 	Our special thank goes to Egor Shevchik from the Joint Institute for Nuclear Research, Dubna, Russia, for the many fruitful discussions on muon vetoes and coupling of WLS fibers to plastic scintillators.
	
	This work has been financed by the CEA, the INFN, the ÖAW and partially supported by the MPI für Physik, by the DFG through the SFB1258 and the Excellence Cluster ORIGINS, and by the European Commission through the ERC-StG2018-804228 "NU-CLEUS". 
	

	\bibliographystyle{JHEP} 
	\bibliography{refs}

\end{document}